\documentclass[journal=jacsat,manuscript=article]{achemso}

\usepackage[version=3]{mhchem} 

\author{Eshan Joshi}
\altaffiliation{MathWorks, Natick, MA 01760}
\author{Samuel Somuyiwa}
\altaffiliation{MathWorks, Natick, MA 01760}
\author{Hossein Z. Jooya}
\altaffiliation{MathWorks, Natick, MA 01760}
\email{hjooya@mathworks.com}
\title[An \textsf{achemso} demo]
  {Graph-Convolutional Deep Learning to Identify Optimized Molecular Configurations}

\abbreviations{DL,GCN,QM7-X}

\begin{document}


\begin{abstract}
Tackling molecular optimization problems using conventional computational methods is challenging, because the determination of the optimized configuration is known to be an NP-hard problem. Recently, there has been increasing interest in applying different deep-learning techniques to benchmark molecular optimization tasks. In this work, we implement a graph-convolutional method to classify molecular structures using the equilibrium and non-equilibrium configurations provided in the QM7-X data set. Atomic forces are encoded in graph vertices and the substantial suppression in the total force magnitude on the atoms in the optimized structure is learned for the graph classification task. We demonstrate the results using two different graph pooling layers and compare their respective performances. 
\end{abstract}

\section{Introduction}
Physical and chemical properties of a molecule are imposed by the three-dimensional (3D) coordinates of its atoms \cite{Mansimov2019}. The task of determining the optimized molecular configuration for small molecules is a vital part of applications such as quantitative structure-activity relationships (QSAR) in cheminformatics and drug discovery \cite{Hawkins2017}. Instrumental techniques such as X-ray crystallography and electron diffraction which are employed in experimental research for imaging molecular 3D structures are usually time-consuming and costly \cite{Mansimov2019}. Tackling this problem using conventional computational methods in chemistry is also challenging, since the determination of optimized configuration is an NP-hard problem. This is because the computational effort for iterative intra-molecular force-minimizing strategies used to tackle this problem scales exponentially with the number of atoms \cite{Deaven1995,Hartke2001}.  

Frequent demonstrations of strong predictive power of Deep Learning in various areas of chemistry \cite{Coote2019} have drawn researchers' attention to harness its abilities to address this fundamental challenge in molecular chemistry \cite{Maziarka2020,Zhou2019,Xu2019,Elton2019}. However, dealing with the non-Euclidean data structures, like irregular 3D positioning of atoms in molecules, compared with the grid-like nature of images requires a generalization of well-established neural models like convolutional networks \cite{Zhang2019}.      
This problem has been addressed by introducing graph-based neural networks \cite{Kipf2017,Bronstein2017,Wu2021,Scarselli2009}. Graphs provide a natural way of describing molecular structures in which vertices correspond to the atoms and edges correspond to chemical bonds. The method of Graph-convolutional networks (GCN), therefore, have recently attracted increasing attention for its promising applications in molecular structures and in predicting their pharmacological activity and chemical reactivity \cite{Kojima2020,Kearnes2016,Sakai2021,Coley2019}. 

One area where GCN-based algorithms can be used is in determining the lowest energy configuration of molecular structures. This is known to be an NP-hard problem \cite{Wille1985}. In such a problem, because the number of candidate local energy minima grows exponentially with the number of atoms, the computational effort scales exponentially with problem size \cite{Genetic_Algorithm_PRL1995}. In practice, searching a significantly rugged energy landscape by performing expensive quantum mechanical simulation methods, like density-functional theory (DFT), is expensive \cite{Message_Passing_2017}. Recently, there has been increasing interest in applying different Deep Learning techniques to benchmark molecular optimization tasks \cite{Zhou2019,He2021,Elton_Review_2019}.

One of the main challenges of maximizing the potential of Deep Learning techniques is the lack of availability of high-quality data \cite{Coote2019}. Implementing this technique to investigate molecular optimization problems requires large amounts of accurate data on both equilibrium and non-equilibrium structures. QM7-X is a recently developed comprehensive data set of 42 physico-chemical properties spanning the chemical space of $\approx$ 4.2 M equilibrium and non-equilibrium structures of small organic molecules with up to seven non-hydrogen (C, N, O, S, Cl) atoms \cite{QM7X}. In this work, we use some of the local (atom-in-a-molecule) properties available in QM7-X data set as node information, and we use the atomic coordinates to construct the edges of the molecular graph. The molecular graphs then are fed to a graph convolutional network for training. Two different GCN designs are proposed by implementing global average pooling \cite{Average_Pooling}, and global max pooling \cite{UNets_2019} layers. Details are provided on how the data is processed throughout these two networks; and the respective performances are compared.  

\section{Data Pre-processing of the QM7-X Data Set}
The QM7-X data set is constructed from 6,950 molecules from the GDB13 database \cite{GDB13}. There are $\approx$ 4.2 million equilibrium and non-equilibrium molecular structures available in QM7-X. Table~(\ref{tab_1}) lists the partial information extracted and stored in MATLAB (.mat) format\cite{MATLAB} for the equilibrium and the most unstable non-equilibrium configuration for each molecule. All molecular structures in this data set are optimized with third-order self-consistent charge density functional tight
binding (DFTB3). In this method, an analytical force equation is derived for a Cartesian coordinate system using the derivative of the total energy with respect to the atomic coordinates  \cite{DFTB3}. Energies are converged to $10^{-6} eV$ and the accuracy of the atomic forces was set to $10^{-4} eV/\AA$. As indicated in Table~(\ref{tab_1}) we extract the energies and forces which are calculated at the (PBE0) level of theory \cite{PBE0} in QM7-X data set. As will be discussed later, the atomic numbers and the atomic forces will be stored at the graph vertices, while the xyz coordinates of the atoms are used to compute the radial distances between the atoms and construct the weighed adjacency matrix of the molecular graph. 

\begin{table}[h]
\caption {\label{tab_1} List of extracted physicochemical properties from the QM7-X dataset. Each property is represented by a symbol (with units and dimension) and can be found in the HDF5 files using the corresponding HDF5 keys. Atomic forces are reported at PBE0 levels of theory. N is the number of atoms in each molecule (N = 4-23 atoms) \cite{QM7X}.}
\begin{tabular}{l@{\quad}l@{\quad}l@{\quad}l@{\quad}lc}
\hline \hline
Symbol & \multicolumn{1}{c}{Property} &  \multicolumn{1}{c}{Unit} &  \multicolumn{1}{c}{Dimension} &  \multicolumn{1}{c}{HDF5 keys}\\
\hline
$Z$ & \multicolumn{1}{c}{Atomic numbers}    &  \multicolumn{1}{c}{-}  & \multicolumn{1}{c}{N} & \multicolumn{1}{c}{’atNUM’} \\
$R$ & \multicolumn{1}{c}{Atomic coordinates} &\multicolumn{1}{c}{$\AA$}   & \multicolumn{1}{c}{3N} & \multicolumn{1}{c}{’atXYZ’}  \\
$E_{PBE0}$ & \multicolumn{1}{c}{PBE0 energy} &\multicolumn{1}{c}{$eV$}   & \multicolumn{1}{c}{1} & \multicolumn{1}{c}{ ’ePBE0’}  \\
$F_{PBE0}$ & \multicolumn{1}{c}{PBE0 atomic forces}    & \multicolumn{1}{c}{$eV/\AA$}  & \multicolumn{1}{c}{3N} & \multicolumn{1}{c}{’pbe0FOR’} \\
\hline \hline
\end{tabular}
\end{table}

\begin{figure}
\includegraphics[width=1\linewidth, height=7cm]{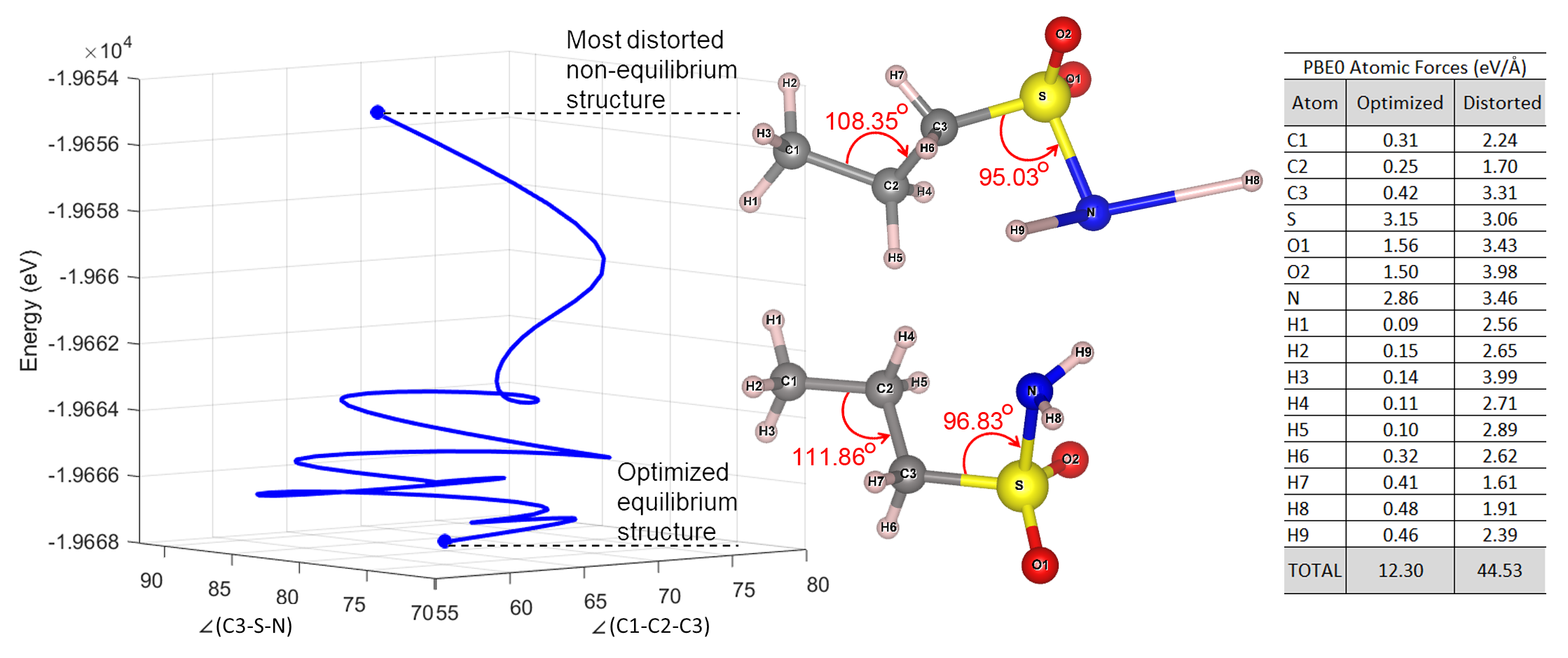}
\caption{\label{fig:Eng_For} The change of the PBE0 molecular energy of propane-1-sulfonamide (C3H9NO2S) with respect to $\angle(C1-C2-C3)$ and $\angle(C3-S-N)$ angles is shown in the 3D plot. The most distorted non-equilibrium configuration (top) and the optimized equilibrium structure (bottom) are shown for comparison. The PBE0 atomic forces for the optimized and most distorted configurations are listed as a table. The total force magnitude is given for each case at the bottom of the table to show the substantially suppressed total force magnitude on the atoms for the optimized structure compared to the most distorted configuration. All the presented data are extracted from QM7-X data set \cite{QM7X}. 
}
\end{figure}

Fig.~\ref{fig:Eng_For} illustrates the molecular structure of Propane-1-sulfonamide (C3H9NO2S) as an example molecule from QM7-X data set. The change of the PBE0 molecular energy with respect to $\angle(C1-C2-C3)$ and $\angle(C3-S-N)$ angles is shown in the 3D plot, where 20 data points are connected by a cubic spline curve using the MATLAB "cscvn" function. The most distorted non-equilibrium configuration (top) and the optimized equilibrium structure (bottom) are shown for comparison. Non-equilibrium structures in QM7-X are generated by displacing each molecular structure along a
linear combination of normal mode coordinates computed at the DFTB3+MBD level \cite{DFTB3_MBD} within the harmonic approximation \cite{QM7X}. The PBE0 atomic forces for the optimized and most distorted configurations are listed as a table in Fig.~\ref{fig:Eng_For}. The prominent feature in this table is the substantially suppressed total force magnitude on the atoms (12.30 $eV/\AA$ for the optimized structure compared to 44.53 $eV/\AA$ for the most distorted configuration). This dramatic force reduction is also seen as a general pattern on individual atoms in this table. Since we are interested in the point of lowest energy, known as a global minimum \cite{Mackay1989}, the forces on some of the atoms may not necessarily get minimized during the optimization process (As an example, see the slightly elevated force magnitude on sulfur atom for the optimized structure in the table, Fig.~\ref{fig:Eng_For}). 

For each of the 6,950 molecules in the QM7-X data set, we extracted only the optimized equilibrium structure and the most distorted non-equilibrium configuration. These are then converted to molecular graphs before being fed to our graph-convolutional network (GCN). For each structure the atomic numbers($Z$) and xyz-coordinates ($R$) are used to generate the adjacency matrix. The binary adjacency matrix for the optimized structure is initially constructed by comparing the radial distances ($r=\sqrt{x^2+y^2+z^2}$) between the atoms, with the available known chemical bond lengths. This binary adjacency matrix is used for the most distorted configuration as well. The same matrix can also be used for each molecule to construct the \textit{weighted} adjacency matrix by replacing the "1"'s with the actual radial distances between the corresponding atoms in both the optimized and most distorted configurations.\\ 
The optimized and distorted graphs for each molecule are stacked in a diagonal block matrix, consisting of 13,900 total molecular data, which will be fed to the graph-convolutional network (GCN) for training. 

\section{Graph Convolutional Network Design}
\begin{figure}
\includegraphics[width=1\textwidth, height=16cm]{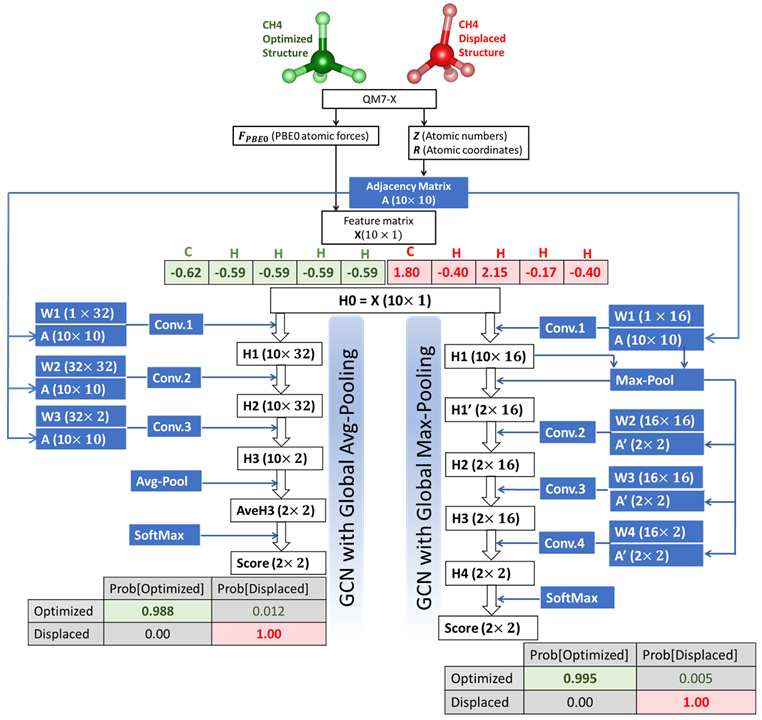}
\caption{\label{fig:Schematics} Schematic illustration of testing the two trained GC networks on optimized and displaced methane molecules. The network with the global average pooling layer \textbf{(left)}, and the network with the global max pooling layer \textbf{(right)} are shown. Both networks start by extracting data from QM7-X data set \cite{QM7X}, and converting it into graph representation. Each convolution layer (indicated by Conv$-l$) uses the trained weight matrix, $W_{l}$, and a relevant adjacency matrix $A$ to update the feature vector $H_{l}$ using Eq.~(\ref{eqn_1}). The initial feature vector $H_{0}=X$ is shown with the magnitudes of the PBE0 atomic forces for the two methane configurations. Note that 3 convolution layers (of size $32$) are used in the left network, compare to the 4 convolution layers (of size $16$) in the right one. As the output scores indicate, both networks are successful in classifying the optimized and displaced configurations.
}
\end{figure}

The GCN models a function $f(X,A)$ on a graph $G=(V,E)$, where $V$ denotes the set of nodes and $E$ denotes the set of edges. The feature matrix $X$ has the dimension $N\times C$, where $N=\mid V \mid$ is the number of nodes in $G$, and $C$ is the number of input channels/features per node. The adjacency matrix $A$ has the dimension of $N \times N$ representing $E$ and describing the structure of $G$. The output of the function $f(X,A)$ is an embedding or feature matrix $H$ of dimension $N \times F$, where $F$ is number of output features per node. In other words, $H$ is the prediction of the network and $F$ is the number of classes. The model $f(X,A)$ is based on spectral graph convolution, with weight/filter parameters shared over all locations in $G$. The model can be represented as a layer-wise propagation model, such that the output of layer $l+1$ is expressed as \cite{Kipf2017}
\begin{equation}\label{eqn_1}
	H_{l+1}=\sigma\left(\hat{D}^{-\frac{1}{2}}\hat{A}\hat{D}^{-\frac{1}{2}}H_{l}W_{l}\right)+H_{l},
    \end{equation}
This GCN model is a variant of the standard GCN model and uses the residual connections between layers (last term on the RHS of the above equation). The residual connections enable the model to carry over information from a previous layer’s input. In this equation, $\sigma$ is an activation function. $H_{l}$ is the activation matrix of layer $l$, with $H_{0}=X$ (magnitudes of the PBE0 atomic forces), and $W_{l}$ is the weight matrix of layer $l$. $\hat{A}=A+I_{N}$ is the adjacency matrix of graph $G$ with added self-connections. $I_{N}$ is the identity matrix. $\hat{D}$ is the degree matrix of $\hat{A}$. The expression $\hat{D}^{-\frac{1}{2}}\hat{A}\hat{D}^{-\frac{1}{2}}$ can be referred to as the normalized adjacency matrix of the graph. \\
Within the above convolutional framework, we have compared the results of implementing global average pooling \cite{Average_Pooling} and global max pooling \cite{UNets_2019} layers, as illustrated in Fig.~\ref{fig:Schematics}. The average pooling layer computes the average of its node features to pool a graph, while the max-node pooling uses the index of the maximum value from each molecular feature to pool the corresponding graph.\\

Fig.~\ref{fig:Schematics} schematically illustrates how these two networks perform when tested on methane molecular configurations, after the weight matrices, $W_{l}$, are trained with the QM7-X data set. The training starts with the weight matrix initialization. In this study, the Glorot method is used to initialize the weight matrices \cite{Glorot_Initializer}. After each epoch, these matrices are updated using the Adam optimization approach \cite{Adam_2015}, until the result is converged. Throughout the network, each convolution layer (indicated by Conv$-l$) uses the weight matrix $W_{l}$ and a relevant adjacency matrix, $A$, as input, to update the feature vector, $H_{l}$, using Eq.~(\ref{eqn_1}). As shown in Fig.~\ref{fig:Schematics}, three convolution layers (of size $32$) are used in the network with global average pooling (left), compare to the four convolution layers (of size $16$) used in the network with the global max pooling (right). The other prominent difference between these two networks is where the pooling layer is used. Having the global average pooling at the end of the convolutions has the benefit of keeping the whole graph structure throughout the network (\textit{i.e.}, the same adjacency matrix is used for all the convolution layers). This, however, is computationally more expensive than what the global max pooling offers. In the latter model, the max pooling happens after the first convolution layer, when the nearest-neighbour effects are included. Reducing the size of the adjacency matrix at this step makes the computations less expensive. The average pooling is particularly justifiable when we deal with small graphs (as is the case in this study with graphs up to 23 nodes). For larger graphs, the max pooling method may be more efficient. As the outcome scores at the end of each network indicate, both of these network designs are successful in classifying the optimized and displaced molecular configurations. 

\begin{figure}
\includegraphics[width=0.9\linewidth, height=16cm]{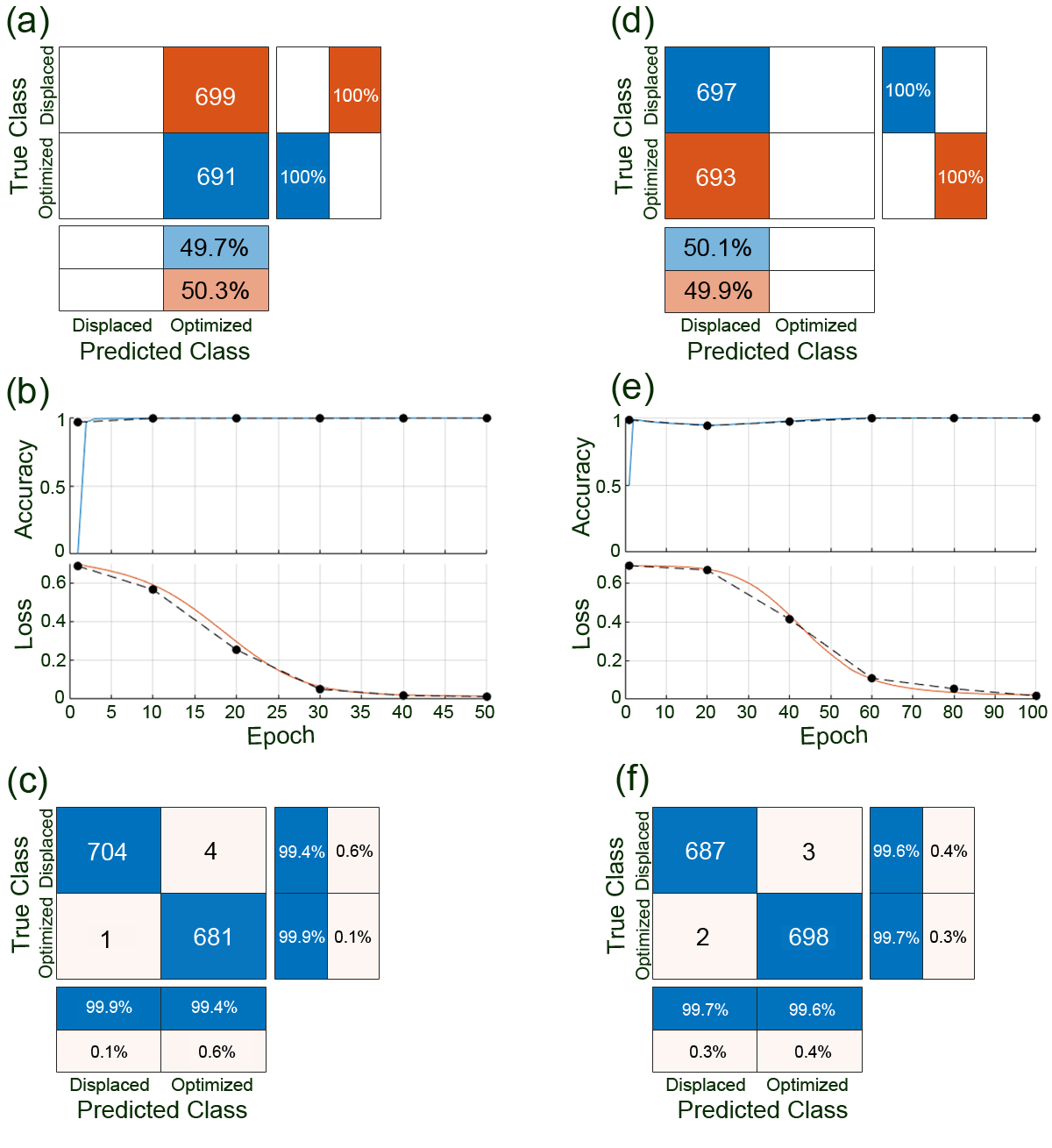}
\caption{\label{fig:GCN-Confusion}  Left panels are the results from the global average pooling network: (a) The confusion matrix obtained from testing the untrained network using random weight matrices, (b) Training progress, and (c) The corresponding confusion matrix obtained from the trained global average pooling network. Right panels, (d-f), are the similar set of results from the global max pooling network. In each confusion matrix, the class-wise precision are the scores in the first row of the 'column summary' of the chart and the class-wise recall are the scores in the first column of the 'row summary' of the chart.}
\end{figure}

Figure ~\ref{fig:GCN-Confusion}(a,d) are the results of testing (using random weights) the untrained GC networks with the global average and the global max pooling networks, respectively. These results indicate that the tuning of the weight matrices via learning is crucial for the performance of these networks. If the architecture of the GC networks itself was essential, then these untrained networks should have been effective for the classification task \cite{Kawamoto_2019}.\\
Figure ~\ref{fig:GCN-Confusion}(b,e) presents the training progress for the global average and the global max pooling networks, respectively. In both cases, the full-batch gradient descent is used. For each epoch, the model gradients and loss are evaluated and the network parameters are updated. The training accuracy score is calculated using an accuracy function which takes the network predictions, the target containing the labels, and the categories classes as inputs and returns the accuracy score. If required, the network is validated by making predictions using the model function and computing the validation loss and the validation accuracy score using cross-entropy and the accuracy function.\\
To visualize how the trained models makes incorrect predictions and evaluate the models based on class-wise precision and class-wise recall, the corresponding confusion matrices are calculated, and visualized in Fig. ~\ref{fig:GCN-Confusion}(c,f). Class-wise precision is the ratio of true positives to total positive predictions for a class. The total positive predictions include the true positives and false positives. A false positive is an outcome where the model incorrectly predicts a class as present in an observation. Class-wise recall, also known as true positive rates, is the ratio of true positives to total positive observations for a class. The total positive observation includes the true positives and false negatives. A false negative is an outcome where the model incorrectly predicts a class as absent in an observation. As demonstrated in this figure, both the global average pooling network, Fig. ~\ref{fig:GCN-Confusion}(c), and the global max pooling network, Fig. ~\ref{fig:GCN-Confusion}(f), successfully identify the optimized molecular configurations from the displaced ones.  

\section{Conclusion}
We propose two architectures for graph convolutional networks to distinguish optimized configurations from displaced molecular structures in the QM7-X data set. Both networks start by encoding PBE0 atomic forces in graph vertices. The substantial suppression in the total force magnitude on the atoms in the optimized structure is learned for the graph classification task. The performance of the designed networks is evaluated when global average pooling is implemented after the last convolution layer, compared to the case when global max pooling is performed after the first convolutional layer. Both architectures show promising results on the QM7-X data set which consists of small organic molecules up to 23 atoms. More experiments are planned to assess the efficiency of these networks when used on larger molecular structures like protein data sets.\\

\begin{suppinfo}
The MATLAB live script code is provided with detailed explanation on each step of the end-to-end workflows for the two proposed GCN architectures.  
\end{suppinfo}

\bibliography{References_DL_GCN}

\end{document}